\documentclass[12pt]{amsart}

\usepackage{graphicx, color}
\usepackage{fullpage}
\usepackage{amssymb, amsthm}
\usepackage{amsmath}
\usepackage{caption}
\usepackage{subcaption}

\newcommand{\be}{\begin{equation}}
\newcommand{\ee}{\end{equation}}
\newcommand{\bc}{\begin{center}}
\newcommand{\ec}{\end{center}}

\begin{document}

\title{\textbf{\large{Non-global parameter estimation using local ensemble Kalman filtering}}}

\author{Thomas Bellsky}
\author{Jesse Berwald}
\author{Lewis Mitchell}
\date{\today}

\begin{abstract}
We study parameter estimation for non-global parameters in a low-dimensional chaotic model using the local ensemble transform Kalman filter (LETKF). 
By modifying existing techniques for using observational data to estimate global parameters, 
we present a methodology whereby spatially-varying parameters can be estimated using observations only within a localized region of space.
Taking a low-dimensional nonlinear chaotic conceptual model for atmospheric dynamics as our numerical testbed,
we show that this parameter estimation methodology accurately estimates parameters which vary in both space and time, 
as well as parameters representing physics absent from the model.
\end{abstract}

\maketitle



\section{Introduction}
Accurately estimating model parameters represents a significant challenge in operational climate modeling.
Model parameters are numerical values typically encoding physical information about a dynamical system:
fundamental constants such as the acceleration due to gravity,
or parameterizations of subgrid-scale processes such as cloud formation \cite{Grabowski2001,Randall2003}.
Modern numerical climate models contain hundreds
of such parameters, all of which must be tuned in some manner so as to ensure 
numerical stability of the code,  
as well as guarantee the best possible match with historic observational data, while also 
providing an adequate parameterization of subgrid-scale physics \cite{Frederiksen2012,Majda2001,Palmer2001}.
Current techniques for estimating such parameters include 
comparing free runs of a model with historical observational data \cite{Peacock2012} 
or Markov Chain Monte Carlo (MCMC) methods \cite{Solonen2012,Urban2010}. 
However, both approaches have limitations: 
the first does not provide a systematic way for using data to inform the model, 
and the second is unsuitable for large models \cite{Annan2007}.

Meanwhile, data assimilation (DA) techniques have been used operationally for more than 20 years to combine a numerical model with observational data in order to provide optimal state estimates for periodic re-initialization \cite{Kalnay2002}.
In recent years,
DA techniques have been used to estimate model parameters in conjunction with updating the state estimate. Typically, there is observable data for the state, but no direct observable data for the parameter. These DA parameter estimation techniques can be divided into two types: the first simultaneously updates the estimated parameter when updating the estimated state (see for example \cite{Anderson2001,Aksoy2006a,Hacker2013}),
and the second separately updates the state estimate and then the parameter estimate \cite{Koyama2010,Ruiz2013}. 
The latter approach utilizes an ensemble Kalman filter (EnKF) to first update an ensemble of state vectors and then separately update an ensemble of global parameters using an EnKF without localization.
A non-localized EnKF is found to be sufficient in this case for estimating global parameters \cite{Koyama2010}. 
\cite{Ruiz2013} modify the above, 
implementing the local ensemble transform Kalman filter (LETKF) to update an ensemble of state vectors and then use an EnKF without localization to again estimate an ensemble of global parameters. 
We note that there are examples of other Kalman filtering-based parameter estimation methods in the literature, for example,
an extended Kalman and ensemble adjustment Kalman filter approach in \cite{Annan2005}
and an EnKF-MCMC hybrid approach in \cite{Hakkarainen2013}.

While the techniques of \cite{Koyama2010} and \cite{Ruiz2013} have been shown to work effectively for estimating global parameters, 
estimating parameters that vary in both space and time remains a significant challenge. In \cite{Annan2005b}, the authors use an augmented ensemble Kalman filter to estimate time-varying parameters for a modified Lorenz-63 model \cite{Palmer1999}. Alternatively, \cite{Kang2011} use the LETKF to estimate a spatially-varying parameter, where they use a technique called ``variable localization,'' which zeros out any covariance between state variables that do not have a physically significant contribution to the estimated parameter.

We propose a robust methodology to estimate both the state vector as well as spatially- and temporally-varying model parameters. 
By only using state observations in a local region of a fixed radius about each grid location,
our method utilizes local information on the state to update the estimated parameter at each grid location. 
We then analyze the fitness of our choice of localization over a range of radii 
to arrive at a localization radius that is optimal for updating a spatially- and temporally-varying parameter. 
Distinct from previous EnKF parameter estimation methods, 
we use the LETKF parameter estimation methodology for both state and parameter estimation, where we investigate a localization scheme in the parameter estimation distinct from the localization scheme in the state estimation. This offers significant improvement in capturing local features of both the state and parameter spaces compared with previous approaches.

We estimate three different types of non-global parameters:
\begin{enumerate}
\item spatially-varying parameters,
\item spatially- and temporally-varying parameters, and
\item parameters representing unobserved physics in the model.
\end{enumerate}

Spatially-varying parameters are a common occurrence in weather and climate prediction.
For example, 
it is well-known that convection statistics in the tropics are very different from those in the mid-latitudes \cite{Xu2001} and so must be parameterized differently in each region. Furthermore, climate models are sensitive to the chosen parameterization of convection \cite{Stainforth2005,Rodwell2007}.
Spatially- and temporally-varying parameters are particularly relevant in climate studies,
for example when estimating or assimilating carbon forcing into models \cite{Engelen2009}. We modify the existing global parameter within our studied conceptual model problem so that it is both a spatially- and temporally-varying parameter.

The use of parameters that represent unobserved physics is a generic
problem across many fields involving prediction. In this case decisions
must be made about how to represent unknown and unobservable processes
in a model.  We will examine two types of parameters representing
unobserved physics that are relevant to weather and climate
prediction. The first type of parameter represents an unobserved
deterministic subsystem evolving on a faster time scale than the
observable slow process \cite{Palmer2001}. The second type of
parameter presents a more difficult challenge. Now, a
stochastic process, whose strength is determined by a deterministic
parameter, forces the true system dynamics, implying that very little
spatial structure can be leveraged by the data assimilation algorithm.

The outline of the remainder of the paper is as follows:
In Section \ref{sec:methodology} we describe the LETKF data assimilation methodology and the proposed parameter estimation methodology,
as well as the conceptual atmospheric model;
in Section \ref{sec:results} we detail results for a conceptual atmospheric model when estimating spatially- and temporally-varying parameters,
as well as for the model error problem when the estimated parameters describe missing physics in the model;
and we conclude with a discussion in Section \ref{sec:discussion}.

\section{Methodology\label{sec:methodology}}
\subsection{Kalman filter data assimilation\label{sec:Kalman}}
The Kalman filter \cite{Kalman} is a recursive state estimation algorithm, which is used to estimate a physical phenomena by combining 
an average of a model forecasting the state with observations of the state. For any positive integer $N$, let $z(t) \in \mathbb{R}^N$ denote the true state vector 
and $M$ be a map that propagates the state vector at time $t_l$ to a new state at time $t_{l+1}$, where
\be
\label{truth} 
z\left(t_{l+1}\right) = M z\left(t_l\right)+\eta.
\ee
Above, $\eta$ is a Gaussian random vector called the model process noise and $Q$ is the covariance matrix for $\eta$, where $\eta \sim N\left(0,Q\right)$. The linear Kalman filter supposes at time step $t_{l+1}$ that we have a previous analysis estimate of the state $z^a\left(t_l\right)$ and a covariance matrix for this analysis state $P^a\left(t_l\right)$, which describes the uncertainty in the analysis state. The linear Kalman filter provides an update of the state estimate at $t_{l+1}$ as the {\em background} solution,
\be
\label{model} 
z^b\left(t_{l+1}\right) = M z^a\left(t_l\right).
\ee
Assuming we are at the $l+1$ time step, the covariance matrix for the background solution is given as
\begin{equation}
\label{Pbmodel} P^b=M P^a M^T + Q.
\end{equation}

Kalman filter techniques suppose there exists a continuous function $H: \mathbb{R}^N \rightarrow \mathbb{R}^m$, where $m$ is also a positive integer and typically $m \ll N$, for which the observations $y \in \mathbb{R}^m$ satisfy $y=H\left(z\right)+\epsilon$; here the observation noise $\epsilon$ is a Gaussian random vector of dimension $m$, where $R$ denotes the covariance matrix for $\epsilon$ such that $\epsilon \sim N\left(0,R\right)$. The linear Kalman filter analysis formulates a maximum likelihood estimate that is found by minimizing the quadratic cost function,
\begin{align}
\label{cost}
 J\left(z\right) =\left(z-z^b\right)^T\left(P^b\right)^{-1}\left(z-z^b\right)  + \left(y-H\left(z\right)\right)^TR^{-1}\left(y-H\left(z\right)\right).
\end{align}
When both the model, $M$, and the observation operator, $H$, are linear, the minimizer of \eqref{cost} is a unique, unbiased, minimum variance estimate of $z$ \cite{LETKF}. The Kalman filter algorithm has many equivalent formulations; we use the following form for the Kalman filter update of the analysis state and the analysis covariance, respectively, at time $t_{l+1}$,
\begin{align}
\label{Kalmanstep} \underbrace{z^a}_\text{analysis} =& \underbrace{ z^b}_\text{background} + \underbrace{K}_\text{Kalman gain} \underbrace{\left(y - Hz^b\right)}_\text{innovation}, \\
\label{analcovar} P^a =& \left(I+P^bH^TR^{-1}H\right)^{-1}P^b.
\end{align}
The Kalman gain matrix $K$ in \eqref{Kalmanstep} is defined as
\begin{equation}
\label{gain} K= P^a H^T R^{-1}.
\ee
We see in \eqref{Kalmanstep} that this update step is a weighted average between the model estimate background state $z^b$ and the observations $y$.

The Kalman filter has been applied to various complex geophysical models \cite{Ghil}. A particular difficulty found in first generation Kalman filters is the problem of estimating large state vectors resulting from more complex models. In order to reduce computational expense, approximations of the background covariance matrix $P^b$ are often made. In 3D-Var and 4D-Var schemes, the background covariance matrix is replaced by an offline constant or slowly time-varying matrix representing typical forecast uncertainties  \cite{corazza}. 

Ensemble Kalman filter (EnKF) methods have been successful \cite{Evensen, Hamill, Houtekamer} in determining online, low-rank approximations of $P^b$ from an ensemble of $k$ model forecasts where $k \ll N$. At a fixed time step $t_{l+1}$, ensemble Kalman filter methods begin with the previous ensemble of $k$ analysis states $\{z_l^{a(i)}:i=1,2,\dots,k\}$. The model $M$, now possibly nonlinear, provides $k$ background states as
\be
\label{ensbackstates}
z_{l+1}^{b(i)} = M z_l^{a(i)},
\ee
for $i=1,2,\dots,k$. Traditional ensemble Kalman filter methods update the model covariance matrix $P^b$ in ensemble space, thus making computations feasible when the ensemble size $k \ll N$ is small. Dropping the time-step notation, these methods produce the following forecast ensemble covariance,
\be
P_{en}^b= \dfrac{1}{k-1}Z^b\left(Z^b\right)^T,  \label{Penb}
\ee
where $Z^b\in \mathbb{R}^{N \times k}$, and the $i$-th column of $Z^b$ is,
\be
Z^{b(i)} = z^{b(i)} - \bar z^{b}. \label{zb}
\ee
Here, $\bar z^b$ is the mean of the ensemble of background states from \eqref{ensbackstates}. Additionally, traditional EnKF methods produce an analysis ensemble covariance,
\be
P_{en}^a= \dfrac{1}{k-1}Z^a\left(Z^a\right)^T, \label{Pena}
\ee
where $Z^a\in \mathbb{R}^{N \times k}$, and the $i$-th column of $Z^a$ is,
\be
Z^{a(i)} = z^{a(i)} - \bar z^a. \label{za}
\ee
Likewise, $\bar z^a$ is the mean of the $k$ analysis states at time $t_{l+1}$.

It is important to note that the $N \times N$ symmetric matrix $P_{en}^b$ is of rank $k-1$, so it is not invertible. It is one-to-one on its column space $S$, however, so $\left(P_{en}^b\right)^{-1}$ is well-defined in this space, which is where the Kalman filter minimization is performed. The ensemble transform Kalman filter (ETKF) determines an appropriate coordinate system for performing the minimization in $S$, where it uses the columns of $Z^b$, described in \eqref{zb}, to span $S$. Since the columns of $Z^b$ are linearly dependent (they sum to zero), these transform methods regard $Z^b$ as a transformation from some $k$ dimensional space $\tilde{S}$ to $S$. Then for  $ w \in \tilde S$, we have $Z^b w \in S$ and the corresponding model state is 
\be
z = \bar z^b + Z^b w. \label{modeltrans}
\ee

If $w$ is a Gaussian random vector where $w \sim N\left(0,(k-1)I\right)$, then the model state in \eqref{modeltrans} has mean $\bar z^b$  with the same covariance matrix as in \eqref{Penb}. From this, ETKF methods motivate a new cost function on $w$ in $\tilde S$. In particular, when $w^a$ is the minimizer of this new cost function, then we have that $\bar z^a = \bar z^b + Z^b w^a$ minimizes the original cost function in Equation \eqref{cost} \cite{Bishop, LETKF}. Thus, ETKF methods take this $\bar z^a$ to be the updated analysis mean.

One concern of ensemble Kalman filter methods is that the ensemble size $k$ is often too small to provide an accurate approximation of the covariance over the entire spatial domain. 
The local ensemble transform Kalman filter (LETKF) \cite{Ott2004} uses spatial localization to help address this concern. 
The LETKF performs each analysis independently at each model grid point using only observations within a prescribed spatial distance. 
This is easily explained using the linear Kalman filter formulation above. At some $i$-th grid point, assume we specify that we use only observations within a radius $d$ of $z_i^b$ in Equation \eqref{model} to update the next analysis. This results in $r$ observations used in the update step, where $r \leq m$ (recall $m$ the total number of state observations). Additionally, the corresponding covariance background matrix will be in $\mathbb R^{{2d+1} \times {2d+1}}$ for a \textit{new} $z^b \in \mathbb R^{2d+1}$ and the observation operator will map $\mathbb R^{2d+1}$ to $\mathbb R^r$. Then the Kalman filter update step results in a \textit{new} $z^a \in \mathbb R^{2d+1}$ and a new analysis covariance matrix in $\mathbb R^{{2d+1} \times {2d+1}}$, where the center term in the \textit{new} $z^a$ is used to update $z_i^a$ (recall the analysis at the $i$-th grid point). Additionally, each of the localized analysis covariance matrices can be combined to construct a global covariance matrix in $\mathbb R^{N \times N}$, where any entry more than distance $d$ from the diagonal will be zero, eliminating spurious artificial correlations between distant positions. 
Thus, localization results in each local analysis determining different linear combinations of ensemble members and the combined global analysis explores a much larger dimensional space than the $k$ ensemble members alone \cite{LETKF}.

\subsection{Ensemble Kalman filter parameter estimation\label{sec:Pestimation}}

Next, we examine how EnKF methods are used to estimate model parameters, where we use $p$ to represent the model parameter state (likewise $p^b$ represents the background model parameter state and $p^a$ represents the analysis model parameter state). We use state augmentation methods \cite{Aksoy2006a,Hacker2013,Koyama2010,Ruiz2013}, where these methods augment an ensemble of background states $\{z^b_i : i =1, 2, \ldots, k\}$ with an ensemble of background parameter states
$\{p^b_j : j =1, 2, \ldots, k'\}$ to produce an ensemble of augmented analysis states $\{(z^a_i,p^a_j): i =1, 2,\ldots, k; j=1,2,\ldots,k' \}$. 
Note that we will use the same number of ensemble members for both state and parameter estimation ($k=k'$), although it is not essential to do so; in fact, for some situations it may be desirable to use more ensemble members for the state estimation than for the parameter estimation \cite{Ruiz2013}. 

The parameter estimation methods we consider can be divided into two types,
which we refer to as \emph{simultaneous} and \emph{separate} methods.
In simultaneous parameter estimation the $N$-dimensional state and $N_p$-dimensional parameter background vectors are augmented to produce a $N+N_p$ length vector $\{(z_j^b,p_j^b): j =1, 2,\ldots, k \}$,
and then an EnKF data assimilation scheme updates the state and parameter from $(z_j^b,p_j^b)$ \cite{Anderson2001,Aksoy2006a,Hacker2013}. 
In separate parameter estimation, first the state is updated from the background state $\{z^b_j : j =1, 2, \ldots, k\}$, and then the parameters are updated in a separate step from the augmented state $\{(z_j^b,p_j^b): j =1, 2,\ldots, k \}$ \cite{Koyama2010,Ruiz2013}.  This augmentation allows the DA method to update the parameter the same as it would any unobserved state variable, by using the background error covariance \cite{Baek}. We will utilize the separate method in this work; although this type is more computationally burdensome (as discussed below), it has been shown to more significantly reduce estimation error \cite{Ruiz2013}.

In the separate parameter estimation approach from \cite{Koyama2010} and \cite{Ruiz2013}, either the EnKF or the LETKF is used to update the state, and then an EnKF without localization is used to update a global parameter $(N_p=1)$. Evidence is also provided that using localization when updating a global parameter offers little benefit in reducing analysis error. 

Our parameter estimation approach is to use the LETKF, thus localization, to both produce an ensemble of analysis states $\{(z_j^a): j =1, 2,\ldots, k \}$ updated from $\{(z_j^b): j =1, 2,\ldots, k \}$ and to produce an ensemble of parameter states  $\{(p_j^a): j =1, 2,\ldots, k \}$ updated from $\{(z_j^b,p_j^b): j =1, 2,\ldots, k \}$. Specifically, the update of this augmented background state produces an ensemble of analysis states each in $\mathbb R^{N+N_p}$, where the final part of each analysis state within $\mathbb R^{N}$ is the update of each ensemble member of parameter states  $\{(p_j^a): j =1, 2,\ldots, k \}$, where each $p_j^a \in \mathbb R^{N_p}$.  

We use this methodology to estimate non-global parameters to determine how localization in the parameter update step will reduce parameter estimation error. Of note, our methodology is more computationally efficient than using an ETKF without localization, since each localized analysis update uses less observations, resulting in less computational complexity. 

Recall the localization radius $d$ for the LETKF. We represent the localization radius used in the analysis update as $d=r_z$ and the localization radius used in the parameter update as $d=r_p$. Thus, at the $i$-th grid point, the LETKF update of $z^b_i$ will only use observations within a radius of $r_z$. Similarly, for the $i$-th spatial location of the augmented state $(z^b, p^b )$, only observations within a radius of $r_p$ are used to update the $i$-th spatial location of the augmented state. 

One key investigation of our research is to determine the optimal value of
the localization radius $r_p$ for the parameter estimation step relative to the system dimension for different types of spatially- and temporally-varying parameters. Accordingly, we vary $r_p$ in our experiments to determine the effect it has on parameter and forecast error. Of note, in this work we focus primarily on the effect of localization on the parameter estimation, hence the localization radius $r_z$ for the state estimation is held constant for similar experiments.

Within the LETKF methodology, the computational time scales (at most) linearly to the size of the state space \cite{LETKF}. If the simultaneous method has a computational time of $t_{sim}$, then {\em ceteris paribus}, the separate method has a computational time of $t_{sep}$, where
\be
t_{sep}=\frac{2N+N_p}{N+N_p}t_{sim},
\ee
where recall $z \in \mathbb R^N$ and $p \in \mathbb R^{N_p}$.
Thus, for a global parameter $(N_p=1),$ the separate method takes less than twice as long as the simultaneous method. When the parameter space is the same as the state space $(N_p=N)$, the separate method only takes $50\%$ more computational time over the simultaneous method. Additionally, the computational time for the separate method is about $50\%$ longer for a parameter that spatially varies over the whole state space $(N_p=N)$ than if the parameter is global $(N_p=1)$.

\subsection{Model problems\label{sec:L96}}
To investigate the efficacy of our LETKF parameter estimation techniques for determining non-global parameters,
we will use a modified version of the well-known Lorenz-96 model \cite{Lorenz1996} with spatially-varying forcing.
The Lorenz-96 model is a system of ODEs that governs the time evolution of $N$ periodic points,
\begin{equation}
\label{Lorenz96eq} \frac{d X_i}{d t} = \left(X_{i+1} +X_{i-2}\right)X_{i-1} - X_i + F, \quad X_{i\pm N} = X_i,\\
\end{equation}
where we fix $N=40$.
Equation \eqref{Lorenz96eq} was formulated to model certain aspects of the atmosphere, where the nonlinear terms model advection and conserve the total energy \cite{Lorenz1996}. The linear terms dissipate the total energy and F represents external forcing, which strongly determines chaotic properties \cite{Lorenz1996}. Within this model we include a spatially-varying forcing term,
\begin{equation}
F := F(i) = F_0+\alpha \sin(2\pi\beta i/N), \quad {i\pm N} = {i},
\label{eqn:F}
\end{equation}
where $\alpha$ determines the amplitude of the forcing and $\beta$ determines the wavelength. 
We choose the above form for the forcing in Equation \eqref{eqn:F} for both simplicity and similarity to a Fourier sine series;
we do not claim that it is representative of any specific atmospheric forcing.
However,
there are numerous examples in the literature of similar modifications to conceptual atmospheric models;
\cite{Palmer1999} modifies the Lorenz-63 by adding a sinusoidal forcing term,
and both \cite{Orrell2003a} and \cite{Mitchell2012d} use modified versions of the Lorenz-96 model as testbeds for experiments in predictability and data assimilation.

\begin{figure}[h]
\begin{center}
\includegraphics[width=.7\columnwidth]{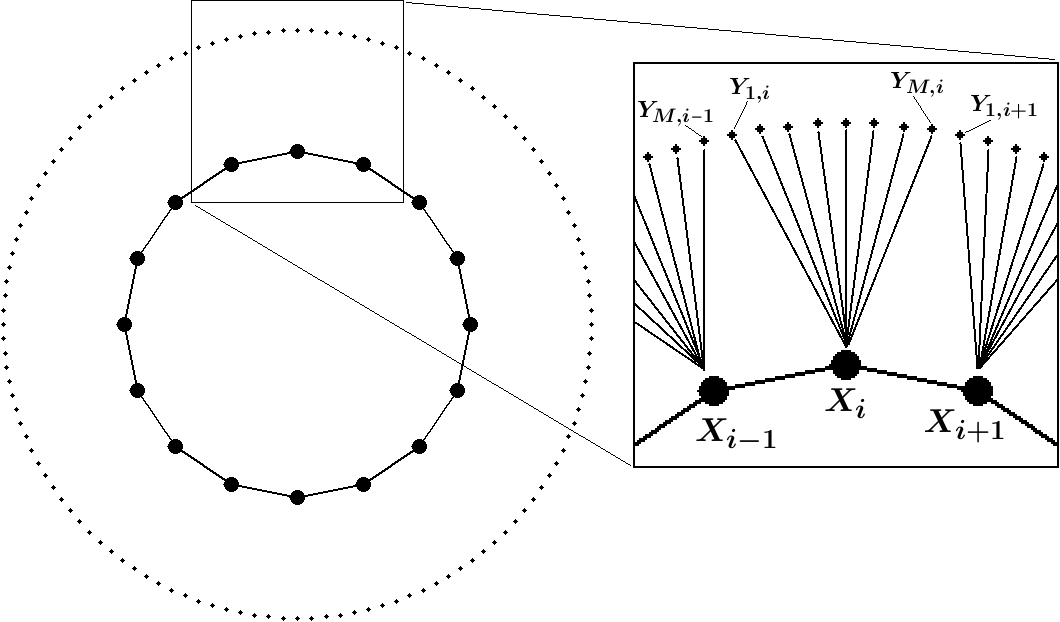}
\caption{Illustration of the coupling topology for the Lorenz-96 fast-slow model for $N=16$ and $M=8$.
\label{fig:L96FS}
}
\end{center}
\end{figure}

To test our techniques on unobserved physics and implicit parameters in atmospheric climatology, we simulate such processes in two ways. 
First, we consider a fast-slow variant of the Lorenz-96 model consisting of $N$ slow variables $X_i$ that are coupled to $M\times N$ fast variables $Y_{j,i},$
\begin{subequations}\label{eqn:fastslow}
\begin{align}
\label{Lorenz96fulleq} \frac{d X_i}{d t} =& \left(X_{i+1} +X_{i-2}\right)X_{i-1} - X_i + F - \frac{hc}{b}\sum_{j=1}^{M}Y_{j,i}, \quad X_{i\pm N} = X_i,\\
\label{Lorenz96fast}	\frac{d Y_{j,i}}{d t} =& cb\left(Y_{j-1,i} -Y_{j+2,i}\right)Y_{j+1,i}-cY_{j,i}+\frac{hc}{b}X_i Y_{M-1,i}, \quad Y_{i\pm M} = Y_i.
\end{align}
\end{subequations}
This model is used to mimic the climatology of the atmosphere, where the fast components $Y_{j,i}$ model weather properties of the atmosphere, which have an 
influence on the slowly evolving components $X_i$ modeling properties of the climate. In order to study a sufficiently chaotic system  \cite{Lieb-Lappen2012}, 
we consider Equation \eqref{eqn:fastslow} with $N=16$ slow components and $M=8$ distinct fast components coupled to each slow component, as illustrated in Figure 
\,\ref{fig:L96FS}. The parameter $h$ determines the level of coupling between the models ($h=0$ leads to no coupling), $c$ determines the time scale 
separation ($c=1$ leads to variables on the same time scale), and $b$ scales the magnitude of the fast components. A common formulation is $h=1$, $c=10$, and 
$b=10$, which induces leading order coupling, where the fast variables $Y_{j,i}$ are $10$ times faster with a magnitude $10$ times smaller than the slow variables $X_i$.

We also consider a model intermediate to the systems defined in \eqref{Lorenz96eq} and \eqref{eqn:fastslow}, which defines a stochastic forcing component to simulate the fast component in~\eqref{Lorenz96fast}. Whereas \eqref{eqn:fastslow} exhibits bidirectional coupling between the subsystems in~\eqref{Lorenz96fulleq} and~\eqref{Lorenz96fast} at different time scales, the stochastic forcing is unidirectional with only the slow component affected by the addition of a stochastic forcing term, simulating stochastic effects of subgrid scale physics. We define each spatial component by
\begin{align}
  \label{eqn:slowstoch}
  d X_i = dt \left[ \left( X_{i+1} + X_{i-2} \right) X_{i} + F(i) \right] + \sigma d W_{t}
\end{align}
where $F(i)$ is defined in Equation ~\eqref{eqn:F} and corresponds to $F_1$ in Table~\ref{tbl:experimentalDesign}. $W_t$ is standard Brownian motion, i.e. a standard Weiner process. For brevity, let $\alpha(X_i(t)) = X_{i+1}(t) + X_{i-2}(t)) X_{i}(t) + F(i)$ and $\beta(t)=\sigma$. Then the standard global Lipschitz condition necessary for {\em strong convergence} of the Euler-Maruyama algorithm,
\begin{align}\label{eqn:lip}
|\alpha(X_{i}(t)) - \alpha(X_{i}(t'))| + | \beta(t) - \beta(t')| \le C |X_{i}(t) - X_{i}(t)|
\end{align}
will be satisfied for some $C>0$. That is, the approximate solution will converge in expectation to the true solution with rate $C\Delta t^{1/2}$ for sufficiently small time step $\Delta t$~(\cite{Higham2003_SDE}). Observe that in terms of the stochastic variable $W_t$ we compute a discretized Brownian path with increments $W(\tau_j) - W(\tau_{j-1})$. A partition of the time interval $[0,T]$ into $\kappa \in \mathbb{N}$ subintervals gives a discrete value for the steps of $\delta t = \tau_{j} - \tau_{j-1} = T/\kappa$. Choosing $\Delta t$ to be an integer multiple of $\delta t$ simplifies the process of the Euler-Maruyama method, guaranteeing that the set of points at which we evaluate the Brownian path is contained in the collection of points $\{t_i\}$ at which the Euler-Maruyama is evaluated.

$W_t$ approximates the influence of coupling unobserved physics
evolving on a fast time scale, and contained in spatially-localized
weather, on climatic changes modeled by the slow variables $X_i$. The
magnitude of the influence due to the unobserved properties are
determined by the deterministic parameter $\sigma$, which corresponds
to the coupling magnitude $b$ in~\eqref{eqn:fastslow}.  In line with
the fast-slow system in~\eqref{eqn:fastslow}, we study a
stochastically-forced system of $N=16$ (slow) variables where the
evolution of each $X_i(t)$ is influenced by a predetermined number of
neighbors, as in Equation~\eqref{Lorenz96eq}, along with an additional
stochastic forcing element of unknown strength. This amounts to
performing data assimilation on a noisy truth for which
spatially-localized physics affect the evolution of the system. It is
worth being precise about the parameter $\sigma$: while we consider
parameter estimation within the context of a stochastic model, the
parameter $\sigma$ that we are estimating is itself deterministic (see
\cite{DelSole2010} and \cite{Yang2009} for more details on the
pitfalls of estimating stochastic parameters).

\begin{figure}[h]
\begin{center}
\includegraphics[width=.7\columnwidth]{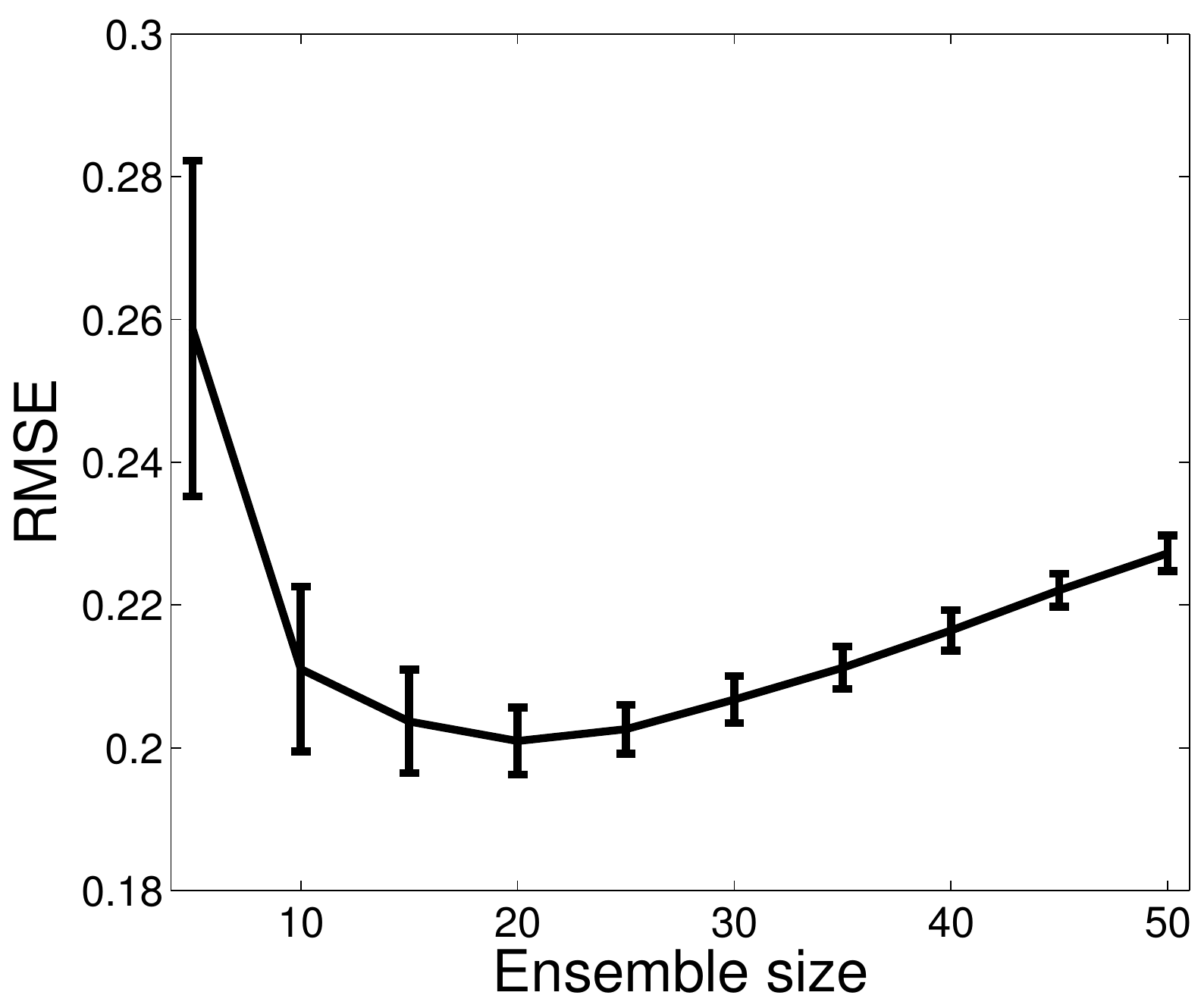}
\caption{RMS errors in the state estimate as a function of ensemble size $k$. We have used an ETKF with no localization or inflation, and estimated a global parameter $F_0 = 8$. Error bars show the standard deviation of the the mean value over 50 independent realizations.
\label{fig:Nens_state}
}
\end{center}
\end{figure}

Table \ref{tbl:experimentalDesign} summarizes our 3 experimental designs.

\begin{table}[t]
\caption{Experimental design for non-global forcing terms.}\label{tbl:experimentalDesign}
\begin{center}
\begin{tabular}{c|l|l}
\hline
 \#  & Experiment type 			& 	Forcing term\\
 \hline \hline
 1 	& Spatially-varying, $F=F(i)$ 	&	$F_1(i) = F_0+\alpha \sin(2\pi\beta i/N)$\\
 2	& Spatially- and temporally-varying, $F = F(i,t)$ & $F_2(i,t) = F_0 + \nu t+\alpha \sin(2\pi\beta i/N)$\\
 3a	& Missing physics, slow-fast deterministic, $F = F(i,j)$	&	$F_{3a}(i,j) = F_1(i) - \frac{hc}{b}\sum_{j=1}^{M}Y_{j,i}$\\
 3b	& Missing physics, stochastic, $F \sim \mathcal{N}(\mu,\sigma)$	& $F_{3b}(i) = F_1(i) + \sigma d W_{t}$\\
\hline
\end{tabular}
\end{center}
\end{table}

\subsection{Error and skill evaluation\label{sec:EandS}}
To evaluate our localized parameter estimation methods when applied to each of the above model problems, we examine the root mean square (RMS) error of both the estimated state and the estimated parameter. We define the RMS error for the LETKF parameter estimation strategy as 
\be
RMSE(t) = \sqrt{\frac{\sum_{i=1}^N \left( Z(x_i,t) -  \bar Z(x_i,t)\right)^2}{N}}, \label{RMSE} 
\ee
where $Z$ is the referenced truth, $\bar Z$ is the mean of the analysis ensemble, and $N$ is the number of grid points. Here, $Z$ can refer to either the state or the parameter. Additionally, we denote the RMS error of the LETKF without localization, that is, the ETKF, as $RMSE_0$. With this, we define the skill of the LETKF parameter estimation strategy as
\be
\label{score} \gamma(t) \equiv \frac{RMSE_0(t)-RMSE(t)}{RMSE_0(t)}.
\ee
Thus, Equation \eqref{score} provides a non-dimensionalized, normalized quantity with which to evaluate our localized methods for each model problem. When $\gamma=1$, the localized scheme perfectly models the true solution (or true parameter), and when $\gamma=0$, the localized scheme provides no improvement over an analysis by the ETKF. 
Note that while this skill measure is bounded above by $\gamma = 1$, 
no such lower bound exists.
In the next section we will use this skill measure to quantify the improvements made by our non-global parameter estimation strategy in determining parameters which vary in both space and time.

\section{Results\label{sec:results}}

\subsection{Spatially-varying forcing\label{sec:space_varying}}

Examining the system in Equation \eqref{Lorenz96eq}, we investigate how well the local parameter estimation methodology estimates a spatially non-uniform but temporally uniform parameter,
that is $F = F(i)$.
To calibrate the ensemble size we first perform $50$ independent experiments where we use an ETKF with no localization or inflation to estimate the state and a global parameter $F = 8$,
and determine the average RMSE in the state estimate as a function of ensemble size. The result of these experiments is in Figure \ref{fig:Nens_state}, which shows a clear minimum RMSE when the number of ensembles is $k=20$.

Next, we calibrate the state estimation inflation and localization parameters $\delta$ and $r_z$ for our experiments on the uncoupled Lorenz-96 system in Equation \eqref{Lorenz96eq} when $N=40$. We perform similar experiments as above for $F=8$, and vary both $\delta$ and $r_z$.
The results for both state and parameter estimation are shown in the contour plots in Figures \ref{fig:inflationLocalization_BW}a and \ref{fig:inflationLocalization_BW}b respectively.
Based on these plots we use a global covariance inflation factor of $\delta = 1.05$ and state localization radius of $r_z = 8$.
Furthermore, 
in all experiments for the uncoupled Lorenz-96 system, we take 20 random observations of the state with Gaussian noise of variance $0.05$, roughly $5\%$ of the climatic variance.

\begin{figure}[h]
\begin{center}
\includegraphics[width=\columnwidth]{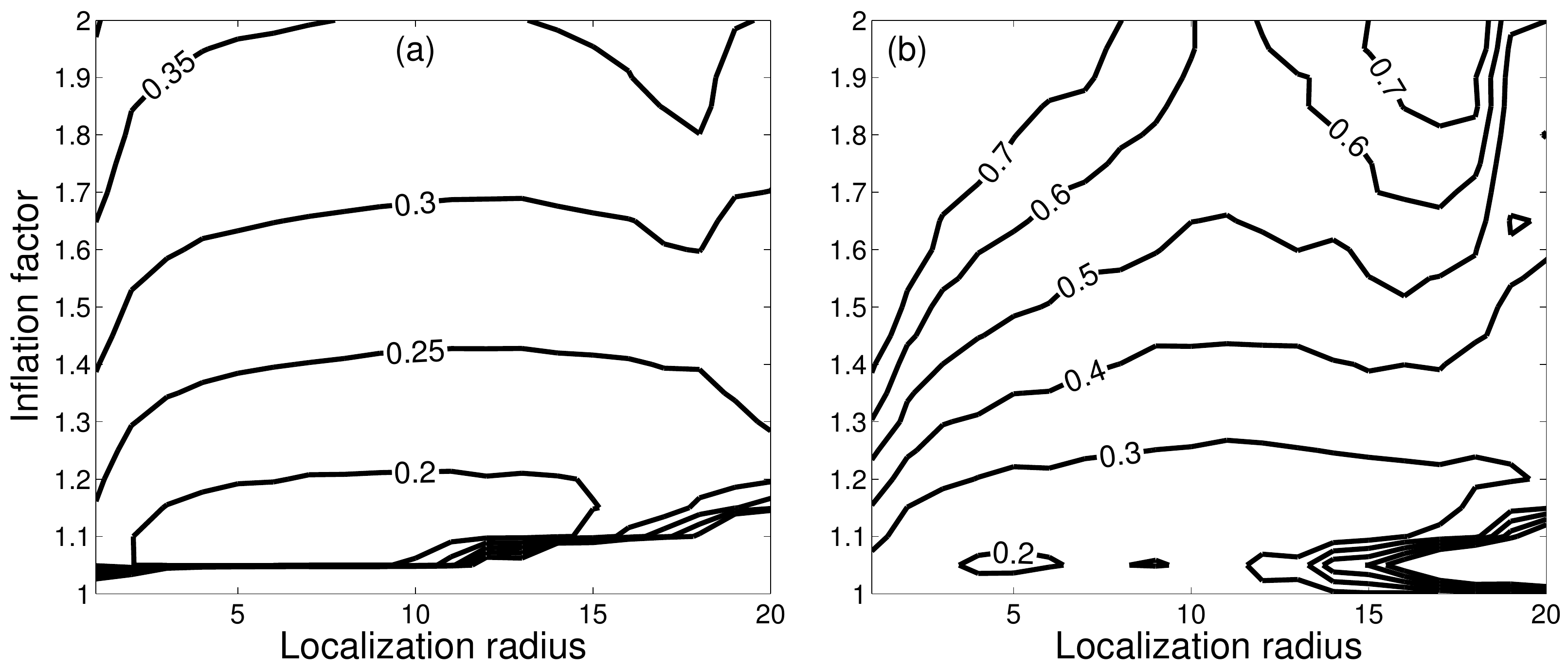}
\caption{Contour plots showing RMS errors as a function of localization radius and multiplicative inflation factor for (a) state estimation and (b) parameter estimation. We have used an ETKF with $k=20$ ensemble members, and the results are averaged over 50 independent realizations.
\label{fig:inflationLocalization_BW}
}
\end{center}
\end{figure}

As a first example of how the parameter estimation performs,
Figure \ref{fig:Fexample} shows examples of the quality of the parameter estimation for a parameter localization radius $r_p = 5,15$ and $20$ (the latter equivalent to estimating the parameters using the ETKF).
This figure gives the final estimate for the spatially-varying forcing  $F(i)$ where forcing amplitude
$\alpha = 2$, 
wavelength
$\beta = 1$,
 and $F_0 = 8$ after 1500 analysis steps,
as averaged over 50 realizations.
The ETKF ($r_p = 20$) approximately recovers the true forcing mean of $F_0 = 8$,
however being a global method, it cannot recover the local spatial variation in $F$.
When $r_p = 15$ the filter is unable to reproduce either the correct mean or spatial variation in $F$ due to spurious long-range correlations between sites.
As Figure \ref{fig:rangeF} shows,
this filter divergence for $r_p=15$ occurs for a range of values of the forcing amplitude $\alpha$ and will be discussed further.
We have checked that the filter divergence is due to finite ensemble size effects;
it does not occur for this value of $r_p$ when using the larger ensemble size of $k=50$.
Finally, we remark that while it appears in Figure~\ref{fig:Fexample} that the average analysis estimate of $F(i)$ is biased towards a lower mean value $F_0$ for $r_p=15$,
the standard deviation of forcing values across realizations is in fact of the same order of magnitude as the estimates themselves.

\begin{figure}[h]
\bc
\includegraphics[width=0.5\columnwidth]{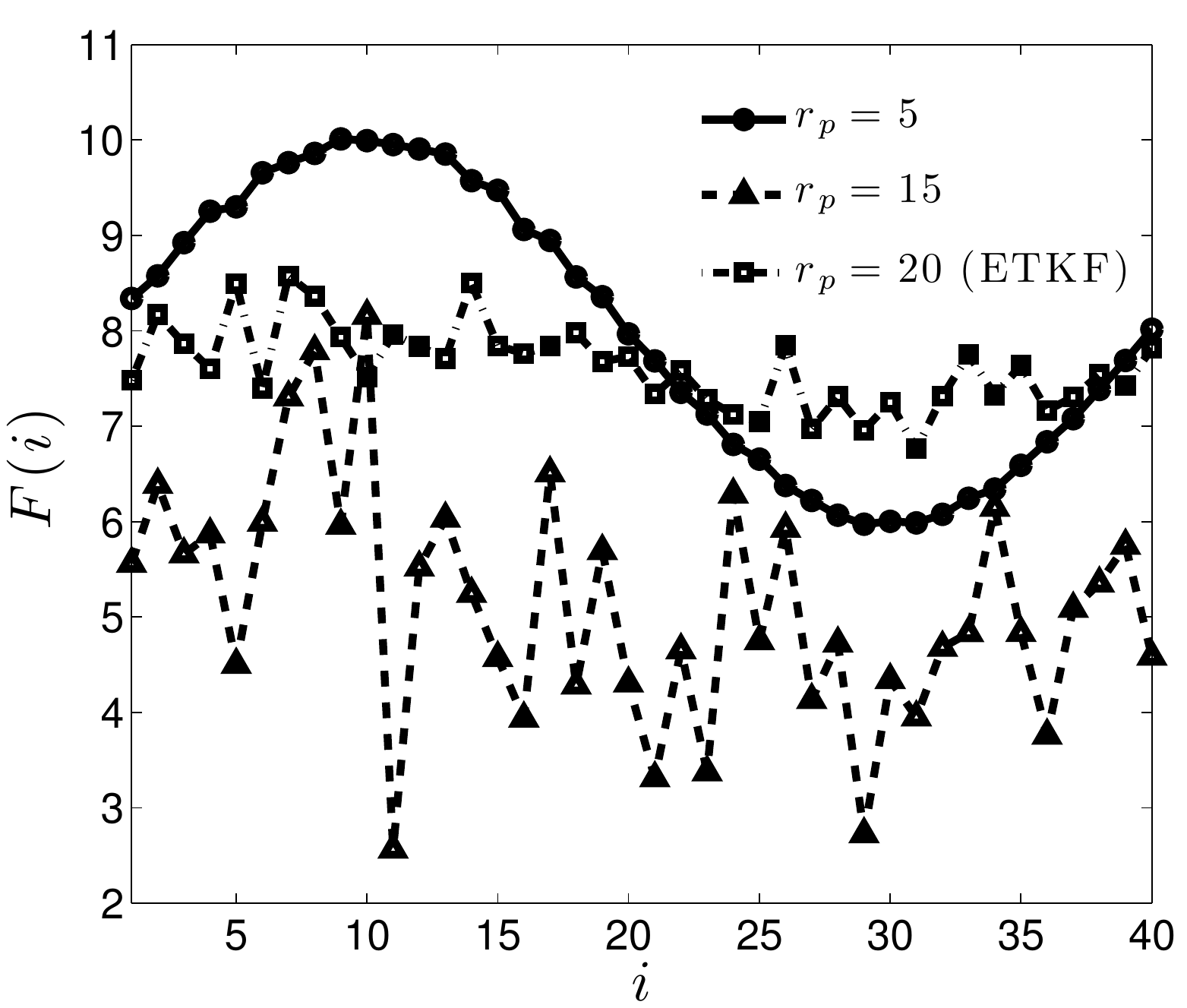}
\caption{Final estimates of $F(i)$ after 1500 analysis cycles for $r_p = 5$ (circles), $15$ (triangles) and $20$ (squares). We have set $\alpha = 2$, $\beta = 1$ and $F_0 = 8$, and averaged over 50 realizations. The true forcing $F(i)$ is visually indistinguishable from the curve with $r_p=5$.
\label{fig:Fexample}
}
\ec
\end{figure}

We now make this comparison of results for different localization radii $r_p$ more clear,
by showing the skill improvement factor $\gamma$ for all possible localization radii $r_p$.
Figure \ref{fig:rangeF} shows how well the localized method performs at estimating the state and non-global forcing parameters as a function of the forcing 
amplitude $\alpha$ and parameter ensemble localization radius $r_p$.
Contours in Figure \ref{fig:rangeF}a show skill improvement for estimates of the state $z$, 
while the contours in Figure \ref{fig:rangeF}b show the skill improvement for estimates of the parameter $F(i)$ in (\ref{eqn:F}).
\begin{figure}[h]
\includegraphics[width=\columnwidth]{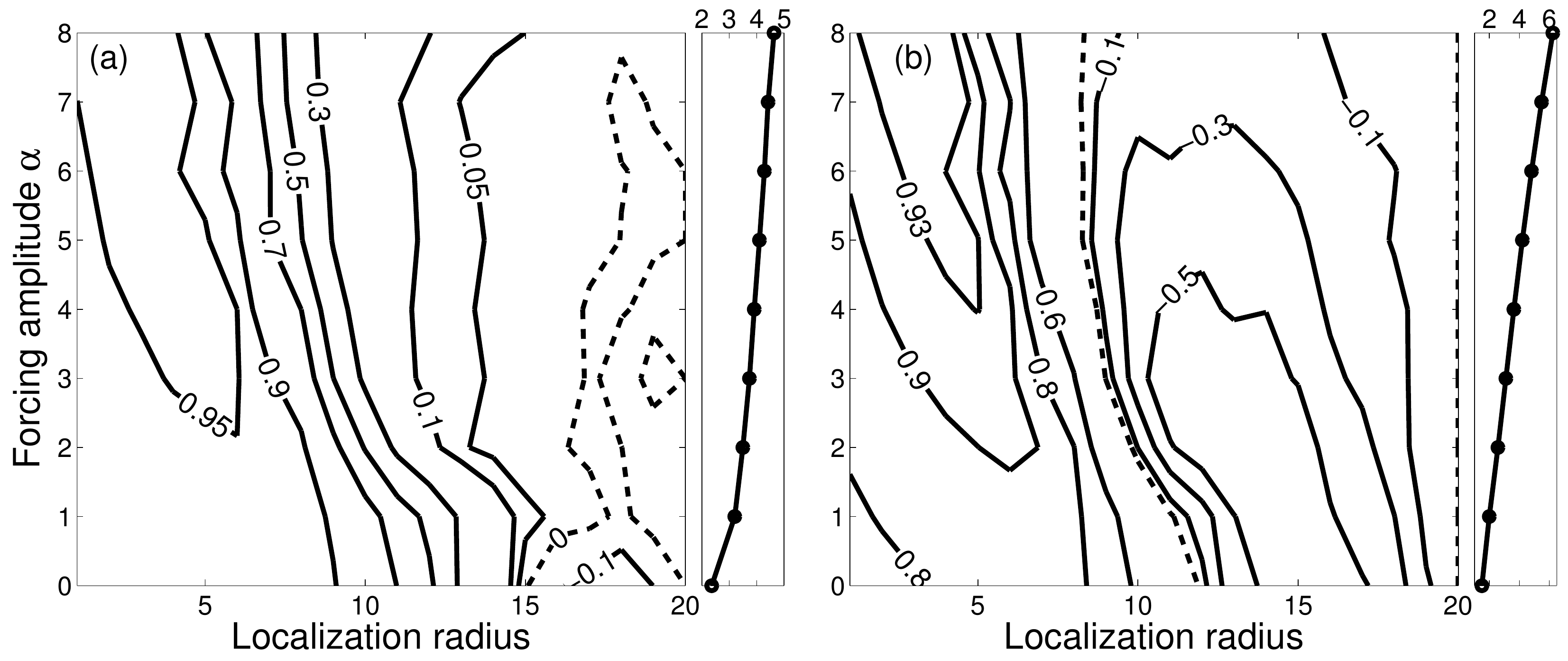}
\caption{
Contours of skill improvement $\gamma$ for (a) state estimation and (b)  parameter estimation as a function of localization radius $r_p$ and parameter amplitude $\alpha$. The insets to the right of each contour plot show the RMS errors for the ETKF, $RMSE_0$, used for the normalization of 
$\gamma$.
The dashed line shows $\gamma = 0$.
\label{fig:rangeF}
}
\end{figure}
The vertical plots to the right of each contour plot show this RMS error of the ETKF as a function of $\alpha$.
As the variation in the forcing increases with increasing amplitude $\alpha$,
progressively smaller local regions must be used in order to obtain accurate estimates of the parameters $F(i)$, these being a substantial improvement over the global ETKF method.
Furthermore, the increase in errors from increasing the localization radius $r_p$ is non-monotonic.
We see a band of localization radii centered around $r_p=14$ where the filter diverges,
and errors are much greater than they are for smaller or larger values of $r_p$.
This band increases in width and shifts to smaller $r_p$ as $\alpha$ increases. 
The effect is more pronounced for the parameter estimation than for the state estimation.

We remark that this filter divergence is a finite ensemble size effect;
increasing the ensemble size to $k=50$ greatly reduces the number of divergences,
however the general trends with $\alpha$ and $r_p$ are preserved.
The range of localization windows over which parameters can be well estimated grows smaller as the parameter amplitude $\alpha$ increases.
This is intuitive, as the variation in forcing $F(i)$ becomes greater as $\alpha$ increases, meaning that only the information coming from observations in increasingly local regions will be useful.
If we instead hold the forcing amplitude constant at $\alpha = 2$ and vary $\beta$ we observe similar trends (not shown) as in Figure \ref{fig:rangeF},
where the skill passes through a local minimum and then a local maximum as $r_p$ increases from 1 up to 20 but little variation in this trend as $\beta$ varies.

\begin{figure}[h]
\includegraphics[width = \columnwidth]{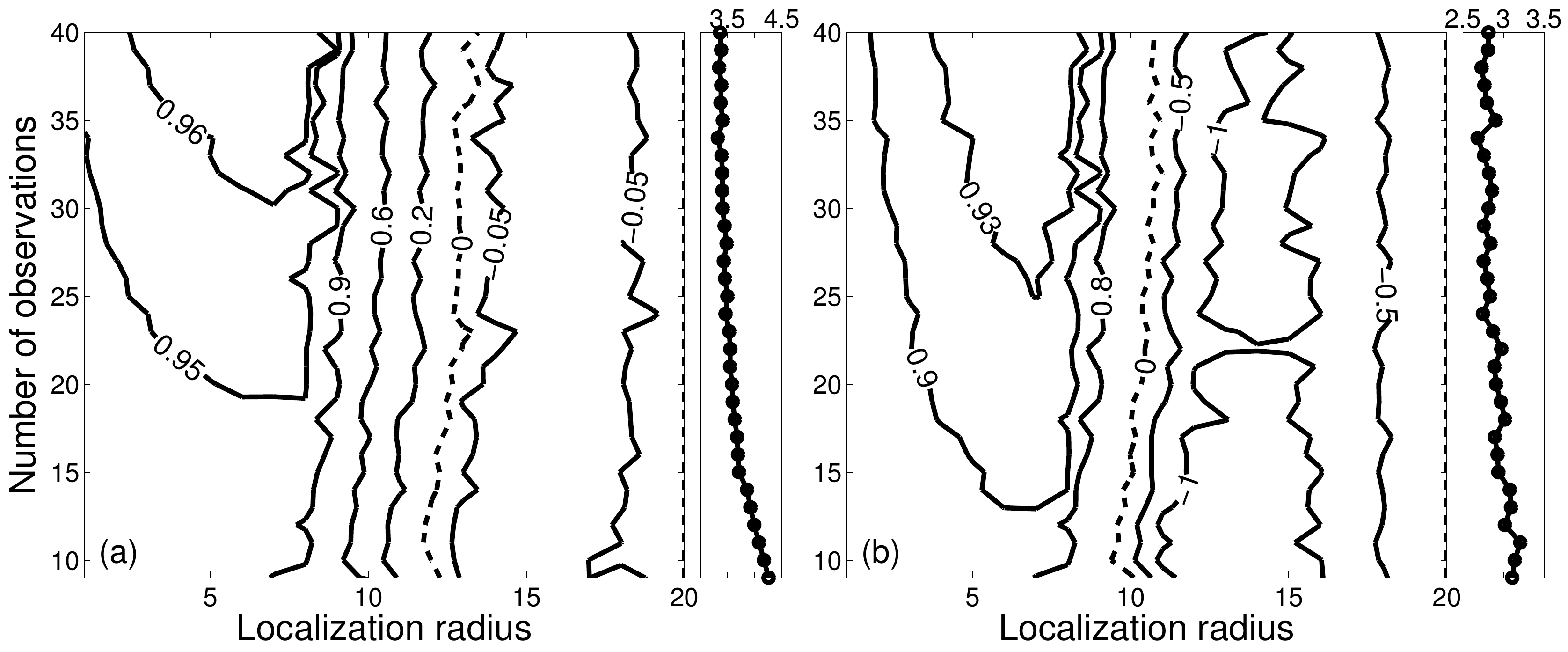}
\caption{Contours of skill $\gamma$ for (a) state estimation and (b)  parameter estimation as a function of both parameter localization radius $r_p$ and number of observations,
for $\alpha = 2$ and
$\beta = 1$.
The insets to the right of each contour plot show the RMS errors for the ETKF, $RMSE_0$, used to normalize $\gamma$.
The dashed line shows $\gamma = 0$.
\label{fig:Nobs}
}
\end{figure}

Finally,
in Figure \ref{fig:Nobs}, for $\alpha = 2$ and $\beta = 1$, we show how the skill depends on the number of observations taken.
Figure \ref{fig:Nobs}a shows contours of skill for state estimates as a function of localization radius $r_p$ and number of observations,
while Figure \ref{fig:Nobs}b shows the same for parameter estimates.
As expected,
when fewer observations are available, it is necessary to use a smaller parameter localization radius $r_p$,
as fewer observations, on average, will be spread further apart in this experiment.
Thus, a larger localization radius will mean that individual sites are informed to a greater degree by distant observations, thereby
corrupting the analysis.
As in Figure \ref{fig:rangeF} there is a band of localization radii for which the filter diverges,
however we remark that there is a slight difference between the state and parameter estimation problems.
Whereas for the state estimation problem the divergence is worst around $r_p = 16$,
for the parameter estimation the divergence is centered around $r_p = 14$.
This could be due to the fact that the observations used are direct measurements of the state,
whereas they only implicitly contain information about the parameters.

\subsection{Spatially- and temporally-varying forcing\label{sec:spacetime_varying}}

Next, again for the system in \eqref{Lorenz96eq}, we investigate how well the LETKF estimates a parameter that varies both spatially and temporally, as described in Table \ref{tbl:experimentalDesign} for the second experiment where $F = F(i,t)$.
Accordingly, we vary (\ref{eqn:F}) so that the mean value $F_0$ increases at a constant rate of 1 forcing unit per 5 units of system time, and set $\alpha = 2$, $\beta = 1$.

Figure \ref{fig:Ft} plots the effect of the localization radius on parameter estimation error, where each figure results from the average of $100$ similar experiments. 
Figure \ref{fig:Ft}a shows time series estimates of $F_1 = F(1,t)$ for various localization radii.
As $F_0$ is continually increasing, all estimates of the forcing $F(i,t)$ eventually diverge from the true value, regardless of the parameter localization $r_p$.
Therefore, in Figure \ref{fig:Ft}b we also show the average time until this divergence occurs as a function of $r_p$,
where the time until divergence is defined as the time when the parameter RMS errors first exceed the threshold value of 3.
The inset to Figure \ref{fig:Ft}b shows the average RMS error for $r_p\leq5$,
showing a clear minimum when $r_p = 2$.

\begin{figure}[h]
\includegraphics[width=\columnwidth]{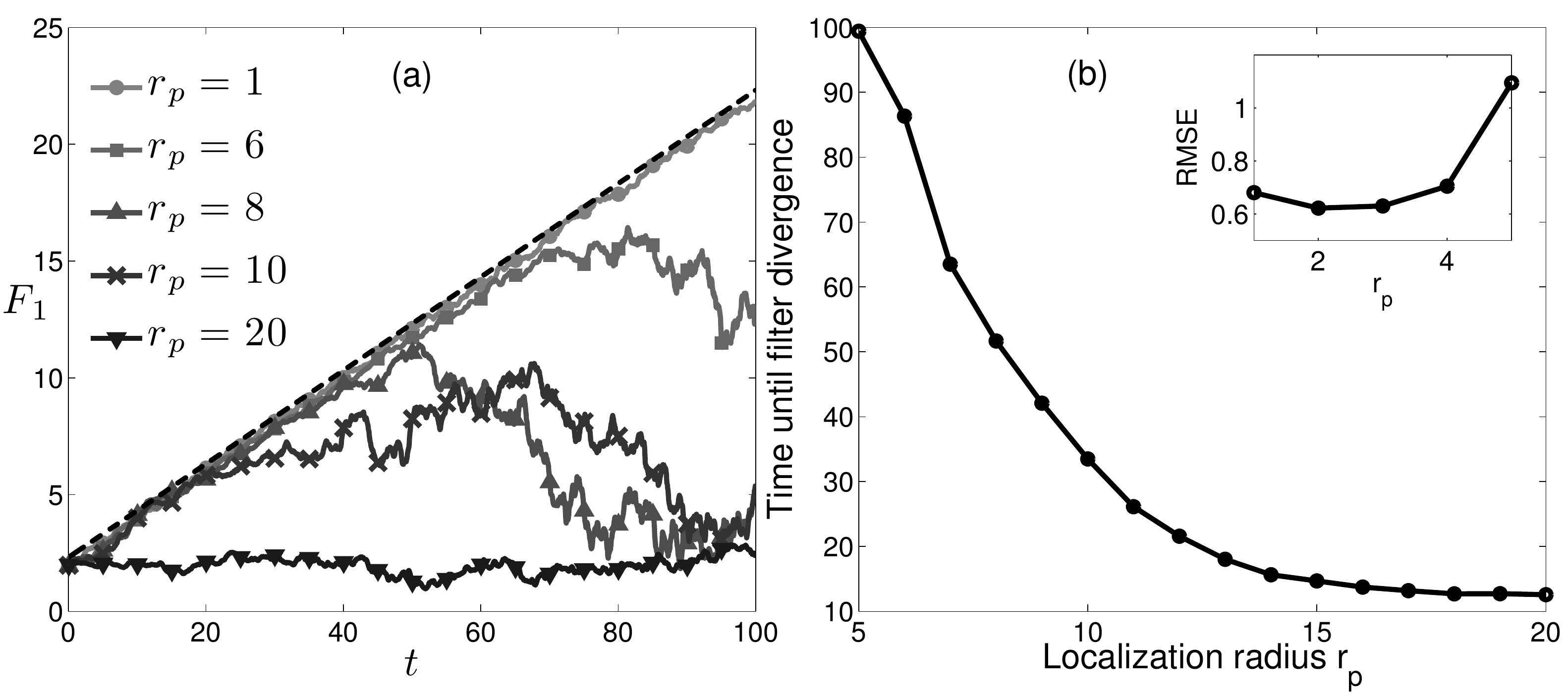}
\caption{Parameter estimation for spatially- and temporally-varying parameter $F = F(i,t)$. (a) The average estimates of $F_1$ for localization radii $r_p=1,6,8,10,20$. The dashed line shows the true parameter value. (b) The average RMS error as a function of localization radius $r_p$.
\label{fig:Ft}
}
\end{figure}

Time series estimates of $F_i$ for $i\ne1$ look qualitatively similar to that in Figure \ref{fig:Ft}a for $F_1$ over the time period shown.
As for the case of the spatially-varying parameter, 
we find that it is necessary to use a small localization window to accurately estimate the parameter over long times.
For $r_p \leq 5$ the analysis estimate of $F(i,t)$ tracks the true parameter value very well,
but diverges from the truth for larger localization windows after increasingly short times.
Increasing the size of the ensemble increases the time until filter divergence for all values of $r_p$,
however we still observe the same trend shown here that filtering with smaller $r_p$ increases the length of fidelity of the experiment.

\subsection{Missing physics problem}\label{sec:fastslow}
All experiments so far have been performed in the perfect model scenario, where the truth is produced by a run of the same model as used in the LETKF forecasting 
step. This is of course overly optimistic -- in an operational situation one is forced to use inadequate models which cannot hope to accurately represent the full 
atmospheric dynamics, either due to coarse resolution or missing physics \cite{Carrassi2011}. Now, we use the uncoupled Lorenz-96 system 
\eqref{Lorenz96eq} to model the true state, where we define the true state to be a solution to the coupled fast-slow Lorenz-96 system \eqref{Lorenz96fulleq}. The model will be 
missing physics from the additional coupled fast term in \eqref{Lorenz96fulleq}, and observations of the truth will include the effects of this additional fast 
term. Thus, when the parameter estimation scheme updates the forcing parameter, it will actually be estimating the forcing parameter plus the additional physics 
from this fast term. 

We again use the local parameter estimation methodology to estimate the spatially-varying forcing parameter described in Table \ref{tbl:experimentalDesign} under experiment 3a where $F_{3a}(i,j) = F_1(i) - \frac{hc}{b}\sum_{j=1}^{M}Y_{j,i}$, with $\alpha=2$ and $\beta=1$. In order to balance low RMS errors and filter stability for the coupled Lorenz-96 system with $N=16$, we have calibrated the ensemble size as $k=20$, the multiplicative inflation factor as $\rho=1.05$, and the localization radius for the state analysis is fixed as $r_z=3.$ This calibration has been done in a similar manner as described in Section \ref{sec:results}\ref{sec:space_varying} and in Figures \,\ref{fig:Nens_state} and \,\ref{fig:inflationLocalization_BW} for the uncoupled Lorenz-96 with $N=40$.

We vary the localization radius for the parameter estimation; recall we have fixed the dimension of the slow variable as $\vec X \in \mathbb R^{16}$, so a localization radius of $r_p=8$ will observe the whole space at every analysis step, thus removing any actual localization and reducing the scheme to the ETKF. We again take randomly located observations of the state at fifty percent of the state space. Hence, there are $8$ observations of $\vec X$ at each time step, where each observation has Gaussian noise of variance $0.05$. 

\begin{figure}[h]
\includegraphics[width=\columnwidth]{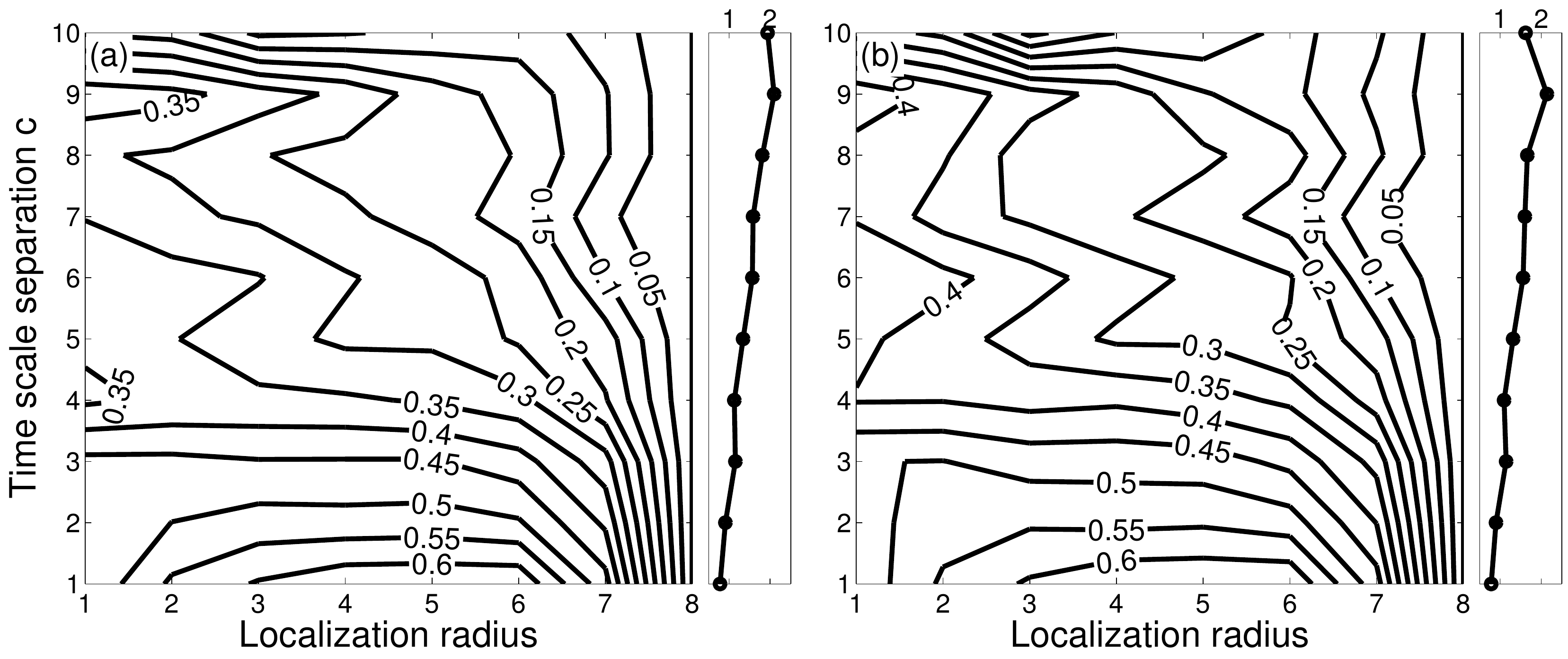}
\caption{Contours of skill improvement $\gamma$ for the (a) estimated state and the (b) estimated forcing. Each is a function of localization radius $r_p$ and time-scale separation factor $c$. The insets to the right of each contour plot show the RMS errors for the ETKF, $RMSE_0$, used for the normalization of $\gamma$.
\label{fig:RMSEvaryc}
}
\end{figure}

Figure \,\ref{fig:RMSEvaryc} 
demonstrates how well the non-global forcing parameter and the state may be estimated as a function of the time-scale separation factor $c$ and the localization radius $r_p$, where we fix the coupling parameter
$h=1$ and $b=10$. 
As in Figures \ref{fig:rangeF} and \ref{fig:Nobs}, Figure \ref{fig:RMSEvaryc}a shows errors in the state estimation,
and Figure \ref{fig:RMSEvaryc}b shows errors in the parameter estimation.
This figure shows skill improvement $\gamma$ contours for estimates of the forcing parameter $F_{3a}(i,j)$, described in Table \ref{tbl:experimentalDesign}, and the resulting state estimates.
In every scenario in Figure \,\ref{fig:RMSEvaryc}, $\gamma >0$, thus the localization strategy offered by the LETKF always outperforms the global ETKF strategy. When estimating the forcing parameter for small time scale separations, there exists a sweet spot for the localization radius $r_p$, where we find that the skill improvement in the estimated forcing is highest for $r_p \in [3,6]$. As seen in both contours, the error is smallest and the skill is highest when the fast and slow components are on the same time-scale. As the time-scale separation becomes larger, the contribution from the fast component to the slow component acts increasingly like noise, making the state and the forcing parameter more difficult to predict. 
Additionally, we see that a smaller localization radius is more optimal as the time scale separation $c$ increases. This is natural for an increased time scale separation, since the contribution of the fast components will behave similarly to noise and the filter will determine decreased slow component structure from observations located further away. 

\begin{figure}[h]
\includegraphics[width=\columnwidth]{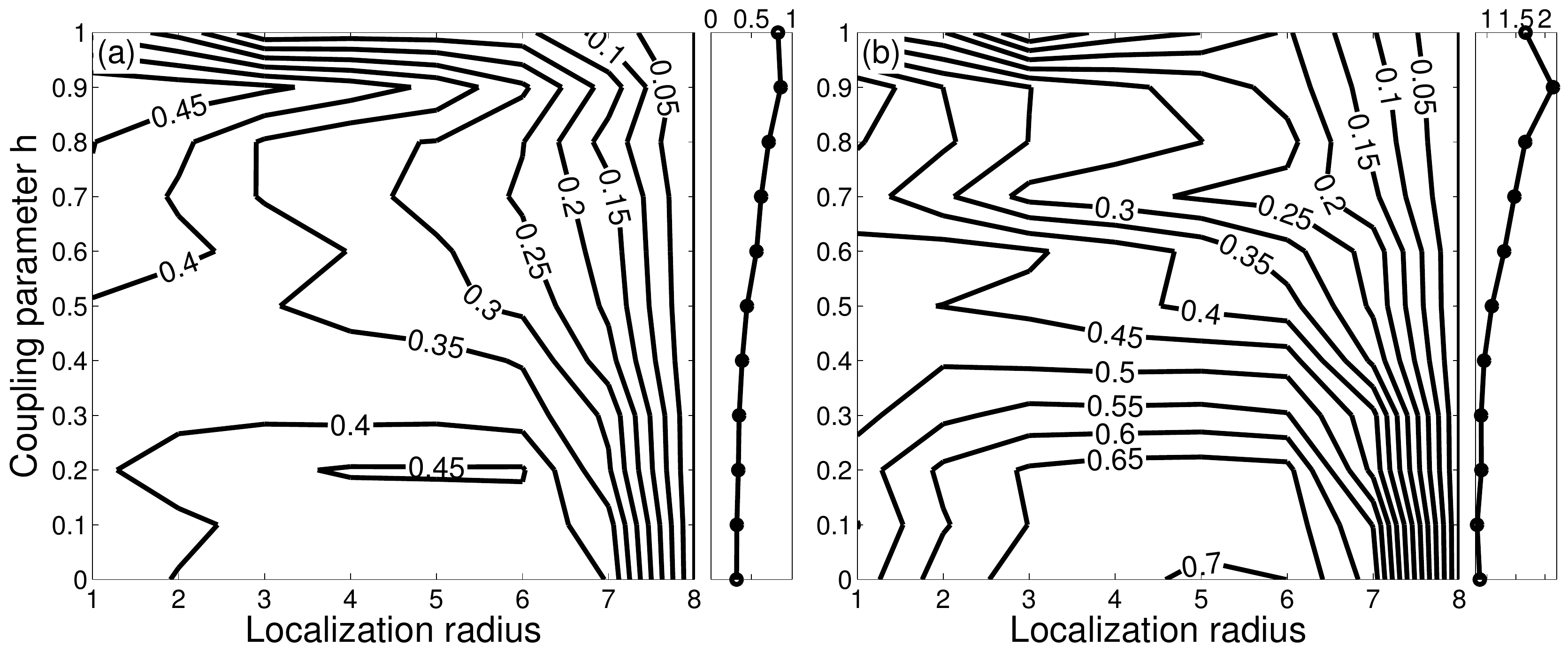}
\caption{Contours of skill improvement $\gamma$ for the (a) estimated state and the (b) estimated forcing, as a function of localization 
radius $r_p$ and coupling parameter $h$. The insets to the right of each contour plot show the RMS errors for the ETKF, $RMSE_0$, used for the normalization of $\gamma$.
\label{fig:RMSEvaryh}
}
\end{figure}

In Figure\,\ref{fig:RMSEvaryh}, we fix $b=10$ and $c=10$ to induce the coupled Lorenz-96 system as explained in Section \ref{sec:methodology}\ref{sec:L96}, where the fast terms are an order of magnitude smaller and an order of magnitude faster than the slow components. We vary the coupling parameter $h$ and parameter ensemble localization radius $r_p$ to determine the optimal localization radius for various levels of coupling. For $h=0$, the contribution from the fast component will always be zero, so in this case the model and the truth are both just the uncoupled Lorenz-96 system as in 
Sections \ref{sec:space_varying} and \ref{sec:spacetime_varying}. As $h$ increases, the contribution from the fast components will increase. 
Figure\,\ref{fig:RMSEvaryh}a shows the skill improvement contour for estimates of the state $\vec X$ and Figure\,\ref{fig:RMSEvaryh}b shows the skill improvement contour for
estimates of the forcing parameter $F_{3a}(i,j) = F_1(i) - \frac{hc}{b}\sum_{j=1}^{M}Y_{j,i}$. As before, the localization scheme always shows improvement over the ETKF. We find that the skill is highest in the estimate of the forcing parameter near $r_p=5$. Additionally, for $r_p=5$, the state estimate has the highest skill when $h=0.2$. In general, skill is higher for smaller localization radii versus larger localization radii, but we see a sweet spot for smaller levels of couplings where the skill is highest approximately for $r_p \in [3,6]$. Again, as expected, the error is smallest when the coupling from the fast component is smallest. As the coupling increases, the contribution from the fast component increases, thus resulting in less accuracy in the estimate of the state
and the estimate of the forcing parameter.  

\subsection{Stochastic model}
\label{sec:stoch}

As described in Section~\ref{sec:methodology}, we approximate the
contribution of the fast component of a fast-slow system by generating
a ``noisy'' truth using Equation~\eqref{eqn:slowstoch}, where now we
are estimating the forcing described in Table
\ref{tbl:experimentalDesign} under experiment 3b as $F_{3b} = F_1(i) +
\sigma d W_{t}$.  The strength of the noise is determined by the
non-stochastic parameter $\sigma$ in Equation~\eqref{eqn:slowstoch},
and is analogous to the coupling parameter, $h$, in
Equation~\eqref{eqn:fastslow}.  We hold the forcing parameters
constant at $\alpha=2$ and $\beta=1$ and vary the localization radius
and magnitude of the stochastic component, $\sigma$.  Thus, as with
the fast-slow system~\eqref{eqn:fastslow}, when we update the
parameters we are simulating the incorporation of unobserved, small
spatial scale physics.

\begin{figure}[h]
\includegraphics[width=\columnwidth]{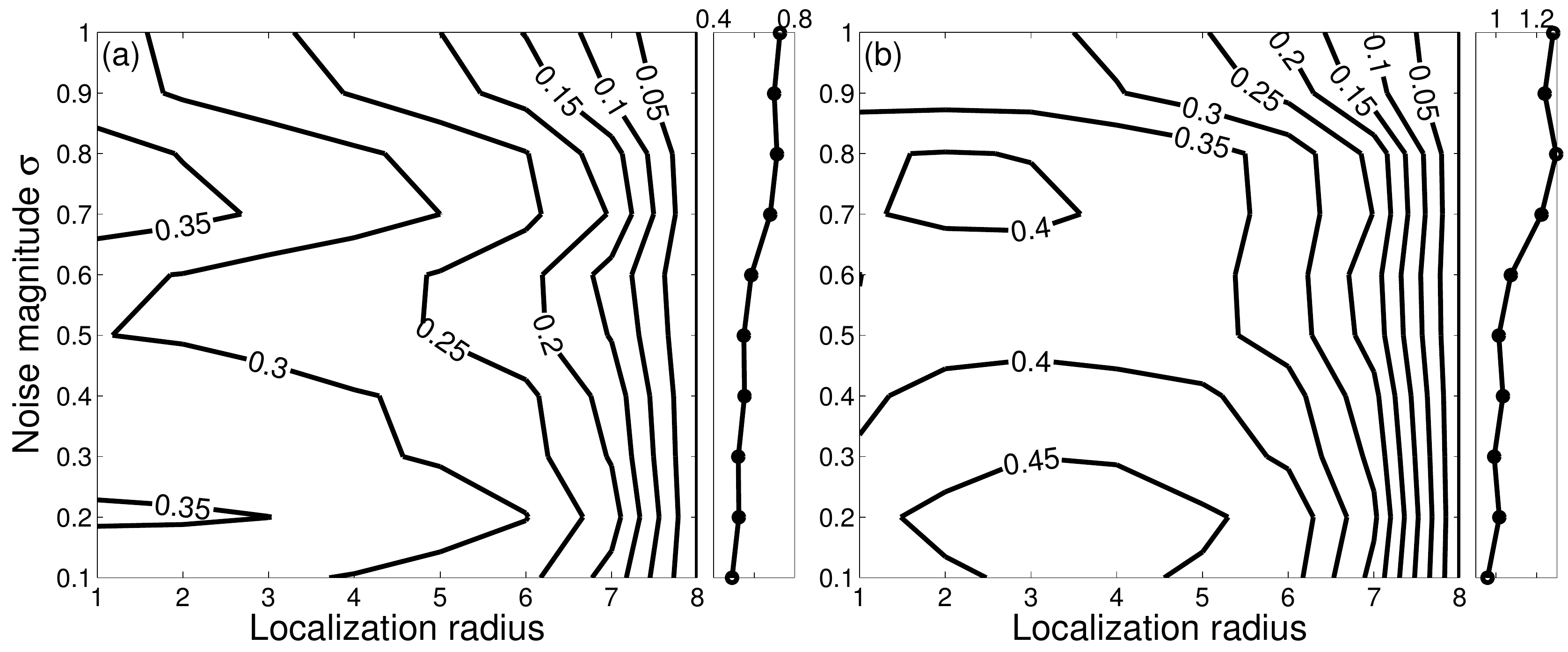}
  \caption{Contours of skill improvement $\gamma$ for the (a) estimated state and the (b) estimated forcing parameter for the stochastic truth defined in Equation~\eqref{eqn:slowstoch}, measured as a function of the localization radius, $r_p$, and the magnitude of the noise, $\sigma$, in the stochastic truth. The insets to the right of each contour plot show the RMS errors for the ETKF, $RMSE_0$, used for the normalization of $\gamma$.}
  \label{fig:stocherror}
\end{figure}

Figure~\ref{fig:stocherror}a shows the skill improvement in the estimated state while Figure~\ref{fig:stocherror}b shows the skill improvement for estimates of the forcing parameter $F(i)$ as the localization radius is increased.  Overall, for a fixed $\sigma$, a smaller localization radius yields more accurate estimates, due to the additive effects of the stochastic component in each $X_i$ in Equation~\eqref{eqn:slowstoch}. But as before, we see a sweet spot in the localization radius $r_p$, where a localization radius spanning about half of the phase space, in particular $r_p \in [2,5]$, is most effective at increasing the skill in the estimation of $F(i)$ for a fixed value of $\sigma$. In this figure, the skill remains relatively constant for a wide range of localization windows as well. Except for fairly high noise levels, the LETKF performs much better than the global ETKF, until larger localization radii above $r_p=6$, where the accuracy of the localization scheme decreases markedly.

In Figure~\ref{fig:stocherror} as the strength of the noise increases, the ability to predict the forcing parameter decreases. This is similar to how increasing the coupling parameter $h$ in the fast-slow system \eqref{eqn:fastslow} leads to more error as seen in Figure \,\ref{fig:RMSEvaryh}. In both experiments $\gamma > 0$ for localization radii $r_p < 8$, showing that the localization technique provides improved estimates compared to the non-local procedure.

\section{Discussion\label{sec:discussion}}

In this article we have presented a methodology for estimating
non-global parameters in numerical models from observational data
using a local ensemble Kalman filtering (LETKF) technique. 
The efficacy of this method is compared to an existing global parameter estimation (ETKF) technique by applying both methods to a low-dimensional conceptual atmospheric system exhibiting chaotic dynamics.
Using this model we have examined the performance of the local parameter estimation method for three different types of non-global parameters,
namely spatially-varying parameters,
spatially- and temporally-varying parameters,
as well as parameters representing unobserved physics. 
For the third set of experiments we have considered two different types of unobserved physics,
provided first by the dynamics of a fully unobserved fast subsystem with
coupling to the slow observed system;
and second by a stochastic process with spatial and temporal correlations.

In most cases the local method is significantly more skillful than the global method at estimating both the parameter and the state,
with a general feature being that the global method will accurately estimate the parameter mean but not spatial deviations from this mean.
The localization radius $r_p$ has been found to be a significant parameter; while there are a wide range of values for which the LETKF method produces improvements over the ETKF method,
the optimal localization radius depends upon spatial characteristics of the parameter being estimated.
In particular, 
smaller localization radii are required as the spatial variation of the parameter is increased,
and as the number of available state observations is decreased. We have found that often there exists a sweet spot for the localization radii, larger than one and less than half the size of the state space, where we find the most skill in parameter estimation.
We have obtained similar results for both model error experiments;
in both cases, 
as the unobserved physics in the nature run becomes noisier
--
either by directly increasing the time scale separation between the slow and the fast systems or the stochasticity
--
it is required to make the parameter localization radius smaller to compensate.

Since this local parameter estimation method performs better than the global method for a low-dimensional conceptual model,
it is of interest to investigate whether the localized method will offer improvement over the non-localized method for a more realistic,
higher-dimensional model. 
In the future, we plan to investigate this by applying these methods
in order to estimate parameters within an intermediate-complexity global circulation model (GCM) such as the simplified parametrization, 
primitive-equation dynamics model SPEEDY \cite{Molteni2003} 
or the weather research forecasting model WRF \cite{Done2004}.
In such studies the shape of the localization region will become a consideration,
as we expect that flow-dependent localization structures will be more desirable than cube-shaped ones.
There may in fact be situations where it could be desirable for the localization radius $r_p$ to vary in space and/or time in response to the local dynamics of the system.
For example, in three dimensions it may become important to use a flow-dependent localization scheme \cite{Bishop2007} to respect the stratified nature of the (relatively flat) atmosphere,
as well as the dynamics of the prevailing flow around orography.
Computationally efficient, 
adaptive diagnostic tools such as bred vectors \cite{Toth1993} or singular vectors \cite{Buizza1995},
as well as the ensemble dimension \cite{Patil2001},
may be useful in such situations.

\section*{acknowledgments} 
This work is supported by the National Science Foundation under the grants DMS-0940363, DMS-0940314, and DMS-0940271.\\
This work was performed in part using computational facilities at the College of William and Mary which were provided with the assistance of the National Science Foundation, the Virginia Port Authority, and Virginia's Commonwealth Technology Research Fund.
LM is grateful for the computational resources provided by the Vermont Advanced Computing Core which is supported by NASA (NNX 08A096G), and the Vermont Complex Systems Center. 

\bibliographystyle{plain}
\bibliography{paramEst2}

\clearpage

\end{document}